\documentclass[12pt,letter]{article}
\pdfoutput=1
\usepackage{graphicx, epsfig, color,cite}
\usepackage{amsmath}
\usepackage{amssymb}
\usepackage{float}
\usepackage{caption,subcaption,graphicx}
\usepackage{hyperref}

\def\tK{\tilde{K}}
\def\hW{\hat{W}}

\def\to{\rightarrow}

\def\bi{\begin{itemize}}
\def\ei{\end{itemize}}

\def\tu{\tilde u}

\def\tf{\tilde f}

\def\tst{\tilde t}

\def\tg{\tilde g}

\def\tw{\widetilde\chi^{\pm}}
\def\tz{\widetilde\chi^0}
\def\alt{\lesssim}
\def\agt{\gtrsim}
\def\be{\begin{equation}}  
\def\ee{\end{equation}}  
\def\bea{\begin{eqnarray}}  
\def\eea{\end{eqnarray}}

\begin{document}
\begin{titlepage}
\begin{flushright}
OU-HEP-200430
\end{flushright}

\vspace{0.5cm}
\begin{center}
{\Large \bf A string landscape guide to soft SUSY breaking terms
}\\ 
\vspace{1.2cm} \renewcommand{\thefootnote}{\fnsymbol{footnote}}
{\large Howard Baer$^1$\footnote[1]{Email: baer@ou.edu },
Vernon Barger$^2$\footnote[2]{Email: barger@pheno.wisc.edu},
Shadman Salam$^1$\footnote[3]{Email: shadman.salam@ou.edu} \\
and
Dibyashree Sengupta$^1$\footnote[4]{Email: Dibyashree.Sengupta-1@ou.edu}
}\\ 
\vspace{1.2cm} \renewcommand{\thefootnote}{\arabic{footnote}}
{\it 
$^1$Homer L. Dodge Department of Physics and Astronomy,
University of Oklahoma, Norman, OK 73019, USA \\[3pt]
}
{\it 
$^2$Department of Physics,
University of Wisconsin, Madison, WI 53706 USA \\[3pt]
}

\end{center}

\vspace{0.5cm}
\begin{abstract}
\noindent
We examine several issues pertaining to statistical predictivity of the string
theory landscape for weak scale supersymmetry (SUSY).
We work within a predictive landscape wherein super-renormalizable terms scan while
renormalizable terms do not. We require stringy naturalness wherein the likelihood of
values for observables is proportional to their frequency within a fertile
patch of landscape including the MSSM as low energy effective theory with a 
pocket-universe value for the weak scale nearby to its measured value in our universe.
In the string theory landscape, it is reasonable that the soft terms enjoy a 
statistical power-law draw to large
values, subject to the existence of atoms as we know them (atomic principle).
We argue that gaugino masses, scalar masses and trilinear
soft terms should each scan independently. 
In addition, the various scalars should scan independently of each other unless
protected by some symmetry.
The expected non-universality of scalar masses-- 
once regarded as an undesirable feature-- 
emerges as an asset within the context of the string landscape picture.
In models such as heterotic compactifications on Calabi-Yau manifolds, 
where the tree-level gauge kinetic function depends only on the dilaton, then
gaugino masses may scale mildly, while scalar masses and A-terms, 
which depend on all the moduli, may scale much more strongly 
leading to a landscape solution to the SUSY flavor and CP problems 
in spite of non-diagonal K\"ahler metrics. 
We present numerical results for Higgs and sparticle mass predictions 
from the landscape within the generalized mirage mediation SUSY model
and discuss resulting consequences for LHC SUSY and WIMP dark matter searches.
\end{abstract}
\end{titlepage}

\section{Introduction}
\label{sec:intro}

The laws of physics as we know them are beset with several fine-tuning problems
that can be interpreted as omissions in our present level of understanding.
It is hoped that these gaps may be filled by explanations requiring additional input 
from physics beyond the Standard Model (SM). 
One of these, the strong CP problem, is solved via the introduction of a 
global Peccei-Quinn (PQ) symmetry and its concomitant axion $a$. 
Another, the gauge hierarchy or Higgs mass problem, 
is solved via the introduction of {\it weak scale supersymmetry} wherein the 
SM Higgs mass quadratic divergences are rendered instead to be more mild log divergences. 
In this latter case, the non-discovery of SUSY particles at LHC has led to concerns
of a Little Hierarchy problem (LHP), wherein one might expect the weak energy scale
$m_{weak}\sim m_{W,Z,h}$ to be in the multi-TeV range rather than 
at its measured value $m_{weak}\simeq 100$ GeV.
A third fine-tuning problem is the cosmological constant (CC) problem, wherein one
expects the cosmological constant $\Lambda\sim m_P^2\sim 6\times 10^{54}$ eV$^2$ 
as opposed to its measured  value $\Lambda\simeq 4.33\times 10^{-66}$ eV$^2$.
The most plausible solution to the CC problem is Weinberg's anthropic solution\cite{Weinberg:1987dv,Weinberg:1987dv2}:
the value of $\Lambda$ ought to be as natural as possible subject to 
generating a pocket universe whose expansion rate is not so rapid that
structure in the form of galaxy condensation should not occur 
(this is called the {\it structure principle}).
 
The anthropic CC solution emerges automatically from the string theory landscape
of (metastable) vacua\cite{Susskind:2003kw} wherein each vacuum solution generates a 
different low energy effective field theory (EFT) and hence 
apparently different laws of physics
(gauge groups, matter content, $\Lambda$, $m_{weak}$ etc.). 
A commonly quoted value for the number of flux vacua in IIB string theory 
is $N_{vac}\sim 10^{500}$\cite{denefdouglas}. 
If the CC is distributed (somewhat) uniformly across its (anthropic) range of values, 
then it may not be surprising that we find ourselves in a pocket universe with
$\Lambda\sim 10^{-120}m_P^2$ since if it were much bigger, we wouldn't be here.
The situation is not dissimilar to the human species finding itself 
fortuitously on a moderate size planet a moderate distance from a stable, 
class-M star: the remaining vast volume of the solar system where we might also 
find ourselves is inhospitable to liquid water and life as we know it and 
we would never have evolved anywhere else.

An essential element to allow Weinberg's reasoning to be predictive 
is that in the subset of pocket universes with varying cosmological constant, 
the remaining laws of physics as encoded in the
Standard Model stay the same: only $\Lambda$ is scanned by the multiverse.
Such a subset ensemble of pocket universes is sometimes referred to as a 
{\it fertile patch}.
Arkani-Hamed {\it et al.}\cite{ArkaniHamed:2005yv} (ADK) argue that only 
super-renormalizable Lagrangian terms should scan in the multiverse 
while renormalizable terms such as gauge and Yukawa couplings 
will have their values fixed by dynamics. 
In the case of an ensemble of SM-like pocket universes with the same gauge group
and matter content, with Higgs potential given by 
\be
V_{SM}=-\mu_{SM}^2\phi^\dagger\phi+\lambda(\phi^\dagger\phi)^2 ,
\ee 
(where $\phi$ is the usual SM Higgs doublet) then just $\mu_{SM}$ and $\Lambda$ should scan. 
This would then allow for the possibility
of an anthropic solution to the gauge hierarchy problem in that the value of
$\mu_{SM}$ (wherein $m_{h}^2(tree)= 2\mu_{SM}^2$) would be anthropically selected to 
cancel off the (regularized) quadratic divergences. Such a scenario is thought to offer
an alternative to the usual application of naturalness, which instead would require
the advent of new physics at or around the weak scale. 

Here, when we refer to naturalness of a physical theory, we refer to 
\begin{quote} 
{\bf practical naturalness:} 
wherein each independent contribution to any physical observable is required to be 
comparable to or less than its measured value. 
\end{quote}
For instance, practical naturalness was successfully used
by Gaillard and Lee to predict the value of the charm quark mass based on contributions to
the measured value of the $K_L-K_S=\Delta m_K$ mass difference\cite{Gaillard:1974hs}.
In addition, it can be claimed that perturbative calculations in theories such as
QED are practically natural (up to some effective theory cutoff $\Lambda_{QED}$).
While divergent contributions to observables appear at higher orders, 
these are dependent quantities: and once dependent quantities are combined, then
higher order contributions to observables are comparable to or less than their
measured values. Thus, we understand the concept of practical naturalness and
the supposed predictivity of a theory to be closely aligned.

To place the concept of naturalness into the context of the landscape of string theory
vacua, Douglas has proposed the notion of {\it stringy naturalness}\cite{Douglas:2004zg}:
\begin{quotation}
{\bf stringy naturalness:} the value of an observable ${\cal O}_2$ 
is more natural than a value ${\cal O}_1$ if more 
{\it phenomenologically viable} vacua lead to  ${\cal O}_2$ than to ${\cal O}_1$.
\end{quotation}
If we apply this definition to the cosmological constant, then 
{\it phenomenologically viable} is interpreted in an anthropic context in that we
must veto vacua which do not allow for structure formation 
(in the form of galaxy condensation). Out of the remaining 
viable vacua, we would expect $\Lambda$ to be nearly as large as 
anthropically possible since there is more volume in parameter space for 
larger values of $\Lambda$. Such reasoning allowed Weinberg to predict the value of
$\Lambda$ to within a factor of a few of its measured value more than 
a decade before its value was determined from experiment\cite{Weinberg:1987dv,Weinberg:1987dv2}.
The stringy naturalness of the cosmological constant is but one example of 
what ADK call {\it living dangerously}: the values of parameters scanned by the
landscape are likely to be selected to be just large enough,  
but not so large as to violate some fragile feature of the world we live in 
(such as in this case the existence of galaxies).

The minimal supersymmetric standard model (MSSM) is touted as a natural solution to
the gauge hierarchy problem. This is because in the MSSM log divergent 
contributions to the weak scale are expected to be comparable to the weak scale
for soft SUSY breaking terms $\sim m_{weak}$. But is the MSSM also more stringy 
natural than the SM? The answer given in Ref. \cite{Baer:2019cae} is {\it yes}.
For the case of the SM valid up to some energy scale $\Lambda_{SM}\gg m_{weak}$,
then there is only an exceedingly tiny (fine-tuned) range of $\mu_{SM}^2$ values which
allow for pocket-universe $m_{weak}^{PU}\sim m_{weak}(measured)$. In contrast, 
within the MSSM there is a very broad range of superpotential $\mu$ values
which allow for $m_{weak}^{PU}\sim m_{weak}(measured)$, provided other
contributions to the weak scale are also comparable to $m_{weak}(measured)$
(as bourne out by Fig's 2 and 3 of Ref.~\cite{Baer:2019cae}). 
For the MSSM, the pocket universe value of the weak scale is given by 
\be
\frac{(m_Z^{PU})^2}{2}=\frac{m_{H_d}^2+\Sigma_d^d-(m_{H_u}^2+\Sigma_u^u)\tan^2\beta}{\tan^2\beta -1}-\mu^2\simeq -m_{H_u}^2-\Sigma_u^u-\mu^2
\label{eq:mzs}
\ee
where the value of $\mu$ is specified by whatever solution to the SUSY $\mu$ problem 
is invoked\cite{Bae:2019dgg}. 
(Here, $m_{H_u}^2$ and $m_{H_d}^2$ are the Higgs field soft squared-masses and the
$\Sigma_d^d$ and $\Sigma_u^u$ contain over 40 loop corrections to the weak scale
(expressions can be found in the Appendix to Ref. \cite{rns})). 
Thus, the pocket universe value for the weak scale is 
determined by the soft SUSY breaking terms and the SUSY-preserving $\mu$ parameter.
If the landscape of string vacua include as low energy effective theories both 
the MSSM and the SM, then far more vacua with a natural SUSY EFT should lead to $m_{weak}^{PU}\sim m_{weak}(measured)$ 
as compared to vacua with the SM EFT where $\Lambda_{SM}\gg m_{weak}$.
In this vein, unnatural SUSY models such as high scale SUSY where 
$m_{soft}\gg m_{weak}$ should be rare occurrences on the landscape as
compared to natural SUSY.

Douglas has also proposed a functional form for the dependence of the 
distribution of string theory vacua on the SUSY breaking scale\cite{Douglas:2004qg}. 
The form expected for gravity/moduli mediation is given by
\be
dN_{vac}(m_{hidden}^2,m_{weak},\Lambda )= f_{SUSY}\cdot f_{EWFT}\cdot f_{CC}\cdot dm_{hidden}^2
\label{eq:dNvac}
\ee
where the hidden sector SUSY breaking scale 
$m_{hidden}^4=\sum_i |F_i|^2 \, +  \, \frac{1}{2}\sum_\alpha D^2_\alpha $
is a mass scale associated with the hidden sector
(and usually in SUGRA-mediated models it is assumed $m_{hidden}\sim 10^{12}$ GeV
such that the gravitino gets a mass $m_{3/2}\sim m_{hidden}^2/m_P$).
Consequently, in gravity-mediation then the visible sector soft terms 
$m_{soft}\sim m_{3/2}$. 
As noted by Susskind\cite{susskind} and Douglas\cite{denefdouglas}, 
the scanning of the cosmological constant is effectively independent 
of the determination of the SUSY breaking scale so that 
$f_{CC}\sim \Lambda /m_{string}^4$. 

Another key observation from examining flux vacua in IIB string theory
is that the SUSY breaking $F_i$ and $D_\alpha$ terms
are likely to be uniformly distributed-- 
in the former case as complex numbers while in the latter case as real numbers.
Then one expects the following distribution of supersymmetry breaking scales
\be
f_{SUSY}(m_{hidden}^2) \, \sim \, (m^2_{hidden})^{2n_F+n_D - 1}
\label{eq:fSUSY}
\ee
where $n_F$ is the number of $F$-breaking fields and $n_D$ is the number of 
$D$-breaking fields in the hidden sector.
Even for the case of just a single $F$-breaking term, then one expects a {\it linear} 
statistical draw towards large soft terms; $f_{SUSY}\sim m_{soft}^n$
where $n=2n_F+n_D-1$ and in this case where $n_F=1$ and $n_D=0$ then $n=1$.
For SUSY breaking contributions from multiple hidden sectors, as typically
expected in string theory, then $n$ can be much larger, with a consequent 
stronger pull towards large soft breaking terms.

An initial guess for $f_{EWFT}$, the (anthropic) fine-tuning factor,
was $m_{weak}^2/m_{soft}^2$ which would penalize soft terms which were much
bigger than the weak scale. This form is roughly suggested by fine-tuning measures
such as $\Delta_{EW}$ where
\be
\Delta_{EW}=|maximal\ contribution\ to\ RHS\ of\ Eq.~\ref{eq:mzs}|/(m_Z^2/2)
\ee
where then $f_{EWFT}\sim \Delta_{EW}^{-1}$.

This ansatz fails on several points\cite{Baer:2017uvn}.
\begin{itemize}
\item Many soft SUSY breaking choices will land one into charge-or-color breaking (CCB) 
minima of the EW scalar potential. 
Such vacua would likely not lead to a livable universe and should be vetoed rather
than penalized.
\item Other choices for soft terms may not even lead to EW symmetry breaking (EWSB).
For instance, if $m_{H_u}^2(\Lambda )$ is too large, 
then it will not be driven negative to trigger spontaneous EWSB. 
These possibilities, labelled as {\it no-EWSB} vacua, should also be vetoed.
\item In the event of appropriate EWSB minima, then sometimes {\it larger} 
high scale soft terms lead to {\it more natural} weak scale soft terms. 
For instance, 1. if $m_{H_u}^2(\Lambda )$ is
large enough that EWSB is {\it barely broken}, then $|m_{H_u}^2(weak)|\sim m_{weak}^2$
(see Fig. 3 of Ref. \cite{Baer:2016lpj}).
Likewise, 2. if the trilinear soft breaking term $A_t$ is big enough, then there 
is large top squark mixing and the $\Sigma_u^u(\tst_{1,2})$ terms enjoy large
cancellations, rendering them $\sim m_{weak}^2$\cite{ltr,rns}. 
The same large $A_t$ values lift the Higgs mass $m_h$ up to the 125 GeV regime.
Also, 3. as first/second generation soft masses are pulled to the tens of TeV regime,
then two-loop RGE effects actually suppress third generation soft terms so that
SUSY may become more natural\cite{Baer:2019zfl}.
\end{itemize}

If one assumes a solution to the SUSY $\mu$ problem\cite{Bae:2019dgg}, 
which fixes the value of $\mu$ so that it can no longer be freely fine-tuned 
to fix $m_Z$ at its measured value, 
then once the remaining SUSY model soft terms are set, one obtains a 
pocket-universe value of the weak scale as an output: 
\textit{e.g} $m_Z^{PU}\ne m_Z(measured)$. 
Based on nuclear physics calculations by Agrawal {\it et al.}\cite{Agrawal:1997gf,Agrawal:1997gf2}, 
a pocket universe value of $m_{weak}^{PU}$ which deviates from our measured value 
by a factor 2-5 is likely to lead to an unlivable universe as we understand it. 
Weak interactions and fusion processes would be highly suppressed and even 
complex nuclei could not form. This would be a violation of the 
{\it atomic principle}: that atoms as we know them seem necessary to support observers.
This is another example of living dangerously: the pull towards large 
soft terms tend to pull the value of $m_{weak}^{PU}$ up in value, but must
stop short of a factor of a few times our measured weak scale lest one jeopardize
the existence of atoms as we know them.
We will adopt a conservative value that the weak scale should not
deviate by more than a factor four from its measured value. This corresponds to a 
value of the fine-tuning measure $\Delta_{EW}\alt 30$.
Thus, for our final form of $f_{EWFT}$ we will adopt\cite{Baer:2017uvn}
\be 
f_{EWFT}=\Theta(30-\Delta_{EW})
\ee
while also vetoing CCB and no-EWSB vacua.

The above Eq. \ref{eq:dNvac} has been used to generate statistical distributions for
Higgs and sparticle masses as expected from the string theory landscape for 
various assumed values of $n=0-4$ and for assumed gravity-mediation model 
NUHM3\cite{Baer:2017uvn} and also for generalized mirage 
mediation\cite{Baer:2019tee}.
For values $n>0$, then there is a statistical pull on $m_h$ to a peak at 
$m_h\simeq 125$ GeV in agreement with the measured value of the Higgs boson mass. 
Also, for $n\ge 1$, then typically the gluino gets pulled to mass values 
$m_{\tg}\sim 4\pm 2$ TeV, {\it i.e.} pulled above LHC mass limits. 
The lighter top squark is pulled to values $m_{\tst_1}\sim 2\pm 1$ TeV 
while higgsinos remain in the $m_{\tz_{1,2},\tw_1}\sim 100-350$ GeV range. 
Since gaugino masses are pulled to large values, 
the neutralino mass gap decreases to $m_{\tz_2}-m_{\tz_1}\sim 3-5$ GeV, 
making higgsino pair production $pp\to \tz_1\tz_2,\ \tw_1\tz_2$ very difficult 
to see at LHC via the soft opposite-sign dilepton signature from 
$\tz_2\to\tz_1\ell^+\ell^-$ decay\cite{Baer:2011ec}. 
Thus, this simple statistical model of the string landscape correctly predicts 
both the mass of the lightest Higgs boson and the fact that LHC sees so far no 
sign of superparticles. And since first/second generation matter scalars are pulled
towards a common upper bound in the $20-40$ TeV range, it also predicts only slight
violations of FCNC and CP-violating processes due to a mixed decoupling/quasi-degeneracy
solution to the SUSY flavor and CP problems\cite{Baer:2019zfl}.

Our goal in this paper is to investigate several issues of soft SUSY
breaking terms relevant for the landscape. The first issue is addressed in 
Sec.~\ref{sec:soft}: which soft terms should scan on the landscape and why. 
The second issue is: of the soft terms which
ought to scan, should they scan with a common exponent $n$, or are there cases where
different soft terms would be drawn more strongly to large values than others:
{\it i.e.} should different $n$ values apply to different soft terms, 
depending on the string model? We address both these issues in Sec. \ref{sec:soft}.
Then in Sec. \ref{sec:results}, we apply what we have learned in Sec. \ref{sec:soft}
to examine how stringy natural are different regions of model parameter space,
as compared to choosing a common exponent $n$ for all scanning soft terms.
Since in the landscape picture many of the soft terms are drawn into the tens-of-TeV
range, we expect a comparable value of gravitino mass $m_{3/2}$, but with TeV scale gauginos.
In such a case, we expect comparable gravity- and anomaly-mediated contributions to soft terms
so that we present our numerical results within the generalized mirage mediation
model GMM$^{\prime}$\cite{Baer:2016hfa}. 
Some discussion on implications for LHC searches along with overall conclusions 
are presented in Sec. \ref{sec:conclude}.

\section{Soft SUSY breaking terms}
\label{sec:soft}

\subsection{Soft terms in the low energy EFT}

In string theory, the starting point is the 10/11 dimensional UV complete string
theory. One then writes the corresponding 10/11 dimensional effective supergravity
(SUGRA) theory by integrating out KK modes and other superheavy states. Compactification
of the 10/11 dimensional SUGRA on a Calabi-Yau manifold (to preserve $N=1$ SUSY in the
ensuing $4-d$ theory) leads to a $4-d$ SUGRA theory containing visible sector fields
plus a plethora (of order hundreds) of gravitationally coupled moduli fields, 
grouped according to complex structure moduli $U_j$ and K\"ahler moduli $T_i$.
In accord with Ref's \cite{Kaplunovsky:1993rd,Brignole:1993dj}, 
we will include the dilaton field $S$ amongst the set of moduli. 
In simple II-B string models, the $U_j$ moduli are stabilized by flux
while the K\"ahler moduli are stabilized by various non-perturbative 
effects\cite{Kachru:2003aw,Balasubramanian:2005zx}. 
In explicit constructions, only one or a few K\"ahler moduli are assumed while realistically 
of order $\sim 100$ may be expected. The moduli stabilization allows in principle
their many vevs to be determined, which then determines the many parameters of the
effective theory. For simplicity, here we will assume the visible sector fields $C_{\alpha}$ 
consist of the usual MSSM fields. 
We will also assume that the moduli $S,\ T_i,\ U_j$ form the hidden sector of the
$4-d$ theory, and provide the required arena for SUSY breaking. 
From this framework, we will then draw conclusions as to precisely which soft terms will scan 
independently within the landscape, and how they are selected for by the power-law
formula $f_{SUSY}\sim m_{soft}^n$.
While many insights into moduli stabilization were made for the case of II-B string theory,
we expect similar mechanisms to occur for other string models (heterotic, etc) since
the various theories are all related by their duality relations.

The $4-d$, $N=1$ supergravity Lagrangian is determined by just two functions that 
depend on the chiral superfields $\phi_M$ of the model: 
the real gauge invariant K\"ahler function
$G(\phi_M,\phi_M^*)=K(\phi_M,\phi_M^*)+\log|W(\phi_M)|^2$ (with $K$ being the real 
valued K\"ahler potential and $W$ the holomorphic superpotental) and the
holomorphic gauge kinetic function $f_a(\phi_M)$. This is presented in units where
the reduced Planck scale $m_P=M_{Pl}/\sqrt{8\pi} =1$. 
The chiral superfields of SUGRA $\phi_M$ are distinguished according to 
visible sector fields $C^\alpha$ and hidden sector fields $h_m$. 
Following \cite{Soni:1983rm,Kaplunovsky:1993rd,Brignole:1993dj}, 
first we expand the superpotential as a power series 
in terms of the visible sector fields:
\be
W=\hat{W}(h_m)+\frac{1}{2}\mu_{\alpha\beta}(h_m) C^\alpha C^\beta +\frac{1}{6}
Y_{\alpha\beta\gamma}(h_m) C^\alpha C^\beta C^\gamma +\cdots
\label{eq:W}
\ee
while the expansion for the K\"ahler potential is
\be
K=\hat{K}(h_m,h_m^*) +\tK_{\bar{\alpha},\beta }(h_m,h_m^*)C^{*\bar{\alpha}}C^\beta +
\left[ \frac{1}{2} Z_{\alpha\beta}(h_m,h_m^*)C^\alpha C^\beta +h.c.\right]+\cdots
\label{eq:K}
\ee
and where the various coefficients of expansion are to-be-determined functions of the
hidden sector fields $h_m$. In the above, Greek indices correspond to visible sector 
fields while lower-case latin indices correspond to hidden sector fields. 
Upper case latin indices correspond to general chiral superfields.

The $F$-part of the scalar potential is given by
\be
V(\phi_M,\phi_M^*)=e^G\left( G_MK^{M\bar{N}}G_{\bar{N}}-3\right)\\
=\left(\bar{F}^{\bar{N}}K_{\bar{N}M}F^M-3e^G\right) 
\label{eq:Vsugra}
\ee

If some of the fields $h_m$ develops vevs such that at least one of the auxiliary fields
$F^m=e^{G/2}\hat{K}^{m\bar{n}} G_{\bar{n}} \ne 0$,
then SUGRA is spontaneously broken. The gravitino gains a mass $m_{3/2}=e^{G/2}$
while soft SUSY breaking terms are generated. 
The soft terms are obtained from the general $4-d$, $N=1$ supergravity Lagrangian\cite{Cremmer:1982en} by replacing the hidden fields $h_m$ and their 
$F_m$-terms by their vevs and then taking the flat limit wherein $m_P\to\infty$ 
while keeping $m_{3/2}$ fixed. 
One is then left with the low energy EFT which consists of a renormalizable 
global SUSY Lagrangian augmented by soft SUSY breaking terms.

The canonically normalized gaugino masses are given by
\be
M_a=\frac{1}{2}(Re\ f_a)^{-1}F^m\partial_m f_a
\label{eq:inos}
\ee

The unnormalized Yukawa couplings are given by
\be
Y^{\prime}_{\alpha\beta\gamma}=\frac{\hat{W}^*}{|\hat{W}|}e^{\hat{K}/2} Y_{\alpha\beta\gamma}
\label{eq:Y}
\ee
while the superpotential $\mu$ terms is given by
\be
\mu^{\prime}_{\alpha\beta}=\frac{\hat{W}^*}{|\hat{W}|}
e^{\hat{K}/2}\mu_{\alpha\beta}+m_{3/2}Z_{\alpha\beta}-
\bar{F}^{\bar{m}}\partial_{\bar{m}}Z_{\alpha\beta}
\label{eq:mu}
\ee

The scalar potential is expanded as
\be
V_{soft}=m^{\prime 2}_{\bar{\alpha}\beta} C^{*\bar{\alpha}}C^\beta +\left(
\frac{1}{6}A^{\prime}_{\alpha\beta\gamma}C^\alpha C^\beta C^\gamma +
\frac{1}{2}B^{\prime}_{\alpha\beta}C^\alpha C^\beta +h.c.\right)
\ee
with {\it unnormalized} soft terms give by
\be
m^{\prime 2}_{\bar{\alpha}\beta}=\left( m_{3/2}^2+V_0\right)\tK_{\bar{\alpha}\beta}-
\bar{F}^{\bar{m}}\left(\partial_{\bar{m}}\partial_n\tK_{\bar{\alpha}\beta}-
\partial_{\bar{m}}\tK_{\bar{\alpha}\gamma}\tK^{\gamma\bar{\delta}}\partial_n\tK_{\bar{\delta}\beta}\right)F^n
\label{eq:m0}
\ee
and
\be
A^{\prime}_{\alpha\beta\gamma}= \frac{\hW^*}{|\hW |}e^{\hat{K}/2}F^m\left[
\hat{K}_mY_{\alpha\beta\gamma}+\partial_m Y_{\alpha\beta\gamma}-
\left(\tK^{\delta\bar{\rho}}\partial_m\tK_{\bar{\rho}\alpha} 
Y_{\delta\beta\gamma}+(\alpha\leftrightarrow\beta)+(\alpha\leftrightarrow\gamma ) \right)
\right] .
\label{eq:A}
\ee
We shall not need the (rather lengthy) expression for $B^{\prime}_{\alpha\beta}$.

\subsection{Implications for the landscape}

\subsubsection{Gaugino masses}

The normalized gaugino mass soft terms are given in Eq. \ref{eq:inos} where
$Re(f_a)=1/g_a^2$. 
For non-zero gaugino masses, then the gauge kinetic function $f_a$ must be a 
non-trivial function of the moduli fields. In most $4-d$ string constructs, then
$f_a$ is taken as $k_a S$ where $k_a$ is the Kac-Moody level of the gauge factor.
This form of the gauge kinetic function leads to universal gaugino masses which
require SUSY breaking in the dilaton field $S$.
The remaining moduli can enter $f_a$ at the loop level and lead to non-universal
gaugino masses. If the moduli-contribution to $M_a$ is comparable to the dilaton
contribution, then one might expect non-universal gaugino masses, but otherwise the
non-universality would be a small effect. If the gaugino masses are dominantly from the
dilaton, then only a single hidden sector field contributes. In this case, one would
expect the $f_{SUSY}$ function to scan as $m_{soft}^1$, {\it i.e. a linear scan} 
for the gaugino masses. In Sec. \ref{sec:results}, we will see that the landscape
actually prefers gaugino masses which are suppressed compared to scalar masses:
$F^S\ll F^m$, where $F^m$ corresponds to the collective SUSY breaking scale
from all the moduli  fields.
In this case, the loop-suppressed moduli-mediated terms may be comparable to the 
dilaton-mediated contribution and non-universality might be expected. Also, even if
moduli-mediated contributions are small, the anomaly-mediated contributions can be 
comparable to the universal contribution. 
To account for this, in Sec. \ref{sec:results} we will work within the 
{\it generalized}\cite{Baer:2016hfa} mirage mediation\cite{Choi:2005ge} scheme for soft term masses, 
and we would indeed expect some substantial non-universality of gaugino masses. 
This type of non-universality leads to gaugino mass unification 
at the mirage scale $\mu_{mir}$ which can be much less 
than the GUT scale $m_{GUT}\simeq 2\times 10^{16}$ GeV where gauge couplings unify. 
Furthermore, in compactification schemes where the moduli-mediated contribution 
to $M_a$ is comparable to the dilaton contribution, then one might expect the 
gaugino masses to scan as $m_{soft}^n$, where the precise value of $n$ 
depends on how many moduli fields contribute to the gaugino masses. 
Since here we are considering that gaugino masses
should scan independently of other soft terms, we will denote their $n$ value
in $f_{SUSY}$ hereafter as $n_{1/2}$.

\subsubsection{Soft scalar masses}

The soft SUSY breaking scalar masses come from Eq. \ref{eq:m0}. In that equation, the
first part, upon normalizing to obtain canonical kinetic terms, leads to 
diagonal and universal scalar masses. In past times, this was a feature to be sought after
since it offered a {\it universality} solution to the SUSY flavor problem\cite{flavor}.
The second term involving partial derivatives of the visible sector K\"ahler metric, 
leads to non-universal soft terms. 
In particular, we would expect non-universal soft scalar masses for the two Higgs
doublets $m_{H_u}^2$ and $m_{H_d}^2$, along with non-universal masses $m_0(1)$, 
$m_0(2)$ and $m_0(3)$ for each of the generations. Intra-generational universality
might be expected to occur for instance where $SO(10)$ gauge symmetry survives 
the compactification (as occurs for instance in some orbifold compactification 
scenarios\cite{Lebedev:2006kn} which lead to local grand 
unification\cite{Buchmuller:2005sh,Ratz:2007my}). 
Then all sixteen fields of each generation which fill out the 16-dimensional 
spinor of $SO(10)$ would have a common mass $m_0(i)$ for $i=1-3$. 
Non-universal soft SUSY breaking scalar masses
lately are a {\it desired} feature in SUGRA models since they allow for radiatively-driven
naturalness (RNS)\cite{ltr,rns}, 
wherein radiative corrections (via RG equations) drive large high
scale soft terms to weak scale values such that the contributions in 
Eq. \ref{eq:mzs} to the weak scale are of natural magnitudes. 
The RNS scenario has a natural home in the string landscape\cite{Baer:2016lpj}. 
For instance, if $m_{H_u}^2$ is statistically favored as large as possible, then instead
of being driven to large, multi-TeV values during the radiative breaking of 
$SU(2)_L\times U(1)_Y$ symmetry\cite{rewsb,rewsb2,rewsb3,rewsb4,rewsb5,rewsb6,rewsb7}, it will be driven to small weak scale
values, just barely breaking EW symmetry. 
This is an example of {\it living dangerously}\cite{ArkaniHamed:2005yv} 
in the string theory landscape, since if the high scale value of $m_{H_u}^2$ 
were much bigger, then EW symmetry wouldn't even break.

As mentioned, the expected non-universality of soft SUSY scalar masses for each generation 
in gravity-mediation was vexing for many years\cite{flavor}, and in fact provided strong 
motivation for flavor-independent mediation schemes such as gauge mediation\cite{gmsb,gmsb2} 
and anomaly-mediation\cite{amsb,amsb2,amsb3,amsb4}.\footnote{A generalized version of AMSB has been 
proposed\cite{namsb} which allows for bulk $A$-terms and non-universal bulk scalar 
masses. This version of AMSB allows for $m_h\sim 125$ GeV and naturalness under the
$\Delta_{EW}$ measure. While winos are still the lightest gauginos, the higgsinos
are the lightest electroweakinos.}
The original incarnations of these models are highly disfavored, if not ruled out, due to
the rather large value of the Higgs mass 
$m_h\simeq 125$ GeV\cite{djouadi,Draper:2011aa,bbm}.
Happily, the string theory landscape offers its own solution to both the SUSY flavor
and CP problems arising from non-universal generations\cite{Baer:2019zfl}. 
In the landscape, the statistical selection of soft SUSY breaking 
scalar masses pulls them to as large of values as possible 
{\it such that their contributions to the weak scale remain of order the weak scale}. 
The top squark contributions to the weak scale are
proportional to the top quark Yukawa couplings, so these soft terms are pulled into 
the few TeV regime. However, first and second generation sfermions have much smaller
Yukawa couplings and so are pulled much higher, into the $20-40$ TeV regime.
In fact, the upper bounds on first/second generation sfermions come from two-loop
RG effects which push third generation soft masses smaller 
(thus aiding naturalness by suppressing $\Sigma_u^u(\tst_{1,2})$ terms) and
then ultimately towards tachyonic. 
From this effect, the anthropic upper bound is the same for both first and second generation
sfermions: they are are pulled to large values, but to a common upper bound. 
This provides a quasi-degenerate, decoupling solution to the SUSY flavor and 
CP problems\cite{Baer:2019zfl}.

Overall, all the SUSY breaking moduli fields should contribute to the soft SUSY breaking 
scalar masses. Thus, we would expect a landscape selection for scalar masses according
to $m_{soft}^{2n_F+n_D-1}$ and thus perhaps a stronger pull on scalar masses to large 
values than might occur for gauginos. To allow for this effect, we hereafter 
denote the value of $n$ contributing to selection of soft scalar masses as $n_0$.

\subsubsection{Trilinears}

The trilinear soft breaking terms, so-called $A$-terms, are given in Eq. \ref{eq:A}.
These terms again receive contributions from all the SUSY breaking moduli fields
and are of order $m_{soft}$. They should scan in the landscape according to 
$f_{SUSY}\sim m_{soft}^{n_0}$, similar to the scalar masses. 
It is worth noting that in Eq.~\ref{eq:A} the Yukawa couplings do not in general factor out of the soft terms.

The statistical selection of large $A$ terms pulls the Higgs mass matrix to maximal
mixing and hence $m_h\to 125$ GeV\cite{ltr,stringy}. 
Meanwhile, it also leads to cancellations in the
loop contributions to the EW scale $\Sigma_u^u(\tst_1)$ and $\Sigma_u^u(\tst_2)$, 
thus {\it decreasing} their contributions to the weak scale. For even larger 
negative values of $A$ parameters, then the $\Sigma_u^u(\tst_{1,2})$ contributions
to $m_{weak}$ increase well beyond $4m_{weak}(measured)$ just before pushing top
squark soft terms tachyonic leading to charge and color breaking (CCB) minima of 
the scalar potential\cite{Baer:2019cae}. 
This is another example of living dangerously.

\subsubsection{$\mu$ parameter}

The bilinear mass term $\frac{1}{2}\mu_{\alpha\beta}(h_m)C^\alpha C^\beta$ in 
Eq.~\ref{eq:W} is forbidden for almost all matter superfields of the MSSM by gauge invariance.
The exception occurs for the vector-like pair of Higgs doublets $\mu H_u H_d$ which contain
opposite hypercharge assignments, making this an allowed term. 
Naively, since the term is supersymmetry preserving, one might expect $\mu\sim m_P$; 
on the other hand, due to the scale invariance of string theory, no mass terms are
allowed for massless states and one gets $\mu=0$\cite{CM}. 
Phenomenologically, such a term with $\mu\sim m_{weak}$ is necessary for 
appropriate EW symmetry breaking. 
The conflict amongst the above issues forms the SUSY $\mu$ problem.\footnote{Twenty 
solutions to the SUSY $\mu$ problem are reviewed in Ref. \cite{Bae:2019dgg}.}
Notice that if $\mu\sim m_{weak}$ in accord with naturalness, but $m_{soft}\agt$ TeV scale,
then $\mu\ll m_{soft}$ and the $\mu$ parameter is also intimately involved 
in the Little Hierarchy (LH) problem: why is there a gap opening up between the weak scale 
and the soft breaking scale? 
The landscape automatically generates such a LH  by pulling soft terms to such large values
that EW symmetry is barely broken.

The analysis of soft SUSY breaking terms already contains within it two possible
resolutions of the $\mu$ problem, which could be acting simultaneously. 
These resolutions depend on the mixing between observable sector fields $H_u$ and $H_d$
with hidden sector fields $h_m$. If a value of $Z_{\alpha\beta}\sim \lambda h_m/m_P$ 
gains a value $\lambda m_{hidden}^2/m_P$ under SUSY breaking, then a 
$\mu$ parameter or order $m_{soft}$ is generated\cite{GM}.

Alternatively, in Eq. \ref{eq:W} where $\mu_{\alpha\beta}$ is a function of 
hidden sector fields $h_m$, 
then if the hidden fields develop a suitable vev, a $\mu$ parameter will be generated.
In the NMSSM\cite{nmssm}, 
a singlet superfield $X$ is added to the visible sector, and when $X$ obtains a 
weak scale vev, then a $\mu$ term is generated. 
If $\mu_{\alpha\beta}$ contains non-renormalizable terms like $\lambda_\mu X^2/m_P$, 
then upon SUSY breaking a $\mu\sim \lambda_\mu m_{hidden}^2/m_P$ is developed with 
$\mu\sim m_{weak}-m_{soft}$. 
This is the Kim-Nilles (KN) mechanism, which originally relied on a PQ symmetry 
to forbid the initial $\mu\sim m_P$ term. An attractive feature of this approach is  that 
the PQ symmetry is also used to solve the strong CP problem via the 
supersymmetrized\cite{sdfsz,sdfsz2,Bae:2013hma} DFSZ axion\cite{dfsz,dfsz2}. 

A less attractive feature is that the global PQ symmetry is not
compatible with gravity/string theory\cite{PQgrav,PQgrav2,PQgrav3,PQgrav4}. A way forward is to invoke instead 
either a (gravity-compatible) discrete gauge symmetry\cite{bgw} $\mathbb{Z}_N$ 
or a discrete $R$-symmetry $\mathbb{Z}_N^R$, where the latter might originate as a 
discrete remnant from 10-d Lorentz symmetry breaking after compactification.
Then the global PQ symmetry emerges as an accidental, approximate symmetry as a consequence
of the underlying discrete gauge or $R$-symmetry. In the latter case, 
a variety of $\mathbb{Z}_N^R$ symmetries have been shown to be anomaly-free and consistent
with grand unification\cite{lrrrssv2} for $N=4,\ 6,\ 8,\ 12$ and $24$. 
The largest of these, $\mathbb{Z}_{24}^R$, 
is strong enough to suppress non-renormalizable contributions to the scalar potential
up to powers of $(1/m_P)^8$, which is enough to solve the strong CP problem 
while maintaining the strong CP angle $\bar{\theta}\alt 10^{-10}$. Such an approach
is attractive since it solves the strong CP problem, solves the SUSY $\mu$ problem,
provides an mechanism for $R$-parity conservation and suppresses otherwise dangerous 
dimension-5 proton decay operators\cite{bbs}.

\section{Results for generalized mirage mediation model 
GMM$^{\prime}$}
\label{sec:results}

\subsection{GMM$^{\prime}$ model and parameter space}

The mirage mediation model is based on comparable moduli- and anomaly-mediated
contributions to soft SUSY breaking terms. 
The boundary conditions are implemented at energy scale $Q=m_{GUT}\simeq 2\times 10^{16}$ GeV where the gauge couplings unify.  
Under this supposition, the gaugino masses receive a universal moduli-mediated
contribution along with an anomaly-mediated contribution which depends on the
gauge group beta functions. The offset from universality is compensated 
for by RGE running to lower mass scales which causes the gaugino masses to 
unify at the {\it mirage} scale $\mu_{mir}=m_{GUT}e^{-8\pi^2/\alpha }$
where $\alpha$ parametrizes the relative moduli- to anomaly-mediated
contributions to the soft terms. 
For $\alpha\to 0$, then one recovers pure AMSB while as $\alpha\to\infty$
then dominant moduli-mediation is recovered. 
The smoking gun signature of mirage mediation is that gaugino masses 
unify at the intermediate mirage scale rather than $m_{GUT}$.
This feature can be tested at $e^+e^-$ colliders operating at 
$\sqrt{s}>2m(higgsino)$\cite{ilc,ilc2}.

Expressions for the soft SUSY breaking terms have
been calculated in 
Ref's \cite{Choi:2005uz,Falkowski:2005ck,Choi:2006xb,Endo:2005uy} under the assumption of simple 
compactifications of II-B string theory with a single K\"ahler modulus.
For more realistic compactifications with many K\"ahler moduli, then
the discrete-valued modular weights are generalized to be continuous 
parameters in the generalized mirage mediation model $GMM^{\prime}$\cite{Baer:2016hfa}
which we adopt here.

For the $GMM^{\prime}$ model, the soft SUSY breaking terms are given by
\begin{eqnarray}
M_a&=& \left( \alpha +b_a g_a^2\right)m_{3/2}/16\pi^2,\label{eq:Ma}\\
A_{\tau}&=& \left( -a_3\alpha +\gamma_{L_3} +\gamma_{H_d} +\gamma_{E_3}\right)m_{3/2}/16\pi^2,\\
A_{b}&=& \left( -a_3\alpha +\gamma_{Q_3} +\gamma_{H_d} +\gamma_{D_3}\right)m_{3/2}/16\pi^2,\\
A_{t}&=& \left( -a_3\alpha +\gamma_{Q_3} +\gamma_{H_u} +\gamma_{U_3}\right)m_{3/2}/16\pi^2,\\
m_i^2(1,2) &=& \left( c_m\alpha^2 +4\alpha \xi_i -\dot{\gamma}_i\right)
(m_{3/2}/16\pi^2)^2 ,\label{eq:mi2} \\
m_j^2(3) &=& \left( c_{m3}\alpha^2 +4\alpha \xi_j -\dot{\gamma}_j\right)
(m_{3/2}/16\pi^2)^2 ,\\
m_{H_u}^2 &=& \left( c_{H_u}\alpha^2 +4\alpha \xi_{H_u} -\dot{\gamma}_{H_u}\right)
(m_{3/2}/16\pi^2)^2 ,\\
m_{H_d}^2 &=& \left( c_{H_d}\alpha^2 +4\alpha \xi_{H_d} -\dot{\gamma}_{H_d}\right)
(m_{3/2}/16\pi^2)^2 .\label{eq:MHd}
\end{eqnarray}
In the above expressions, the index $i$ runs over first/second 
generation MSSM scalars
$i=Q_{1,2},U_{1,2},D_{1,2},L_{1,2}$ and $E_{1,2}$ while $j$ runs overs
third generation scalars $j=Q_3,U_3,D_3,L_3$ and $E_3$. 
Here, we adopt an independent value $c_m$ for the first two
matter-scalar generations whilst the parameter $c_{m3}$ applies to
third generation matter scalars. 
The independent values of $c_{H_u}$ and $c_{H_d}$, which set the
moduli-mediated contribution to the Higgs mass-squared soft terms, may
conveniently be traded for weak scale values of $\mu$ and $m_A$ as is
done in the two-parameter non-universal Higgs model (NUHM2)\cite{nuhm2,nuhm22,nuhm23,nuhm24,nuhm25,nuhm26}.
This procedure allows for more direct exploration of 
stringy natural SUSY parameter space where most landscape solutions 
require $\mu\sim 100-360$ GeV in anthropically-allowed pocket 
universes\cite{Baer:2019cae}.
Thus, the $GMM^{\prime}$ parameter space is given by
\be
\alpha,\ m_{3/2},\ c_m,\ c_{m3},\ a_3,\ \tan\beta ,\ \mu ,\ m_A
\ \ \ (GMM^{\prime} ). 
\label{eq:gmmp}  
\ee
The natural GMM and GMM$^\prime$ models have been incorporated 
into the event generator program Isajet 7.88\cite{isajet} 
which we use here for spectra generation.
(The GMM and GMM$^{\prime}$ models are equivalent: 
GMM uses high scale Higgs soft terms $m_{H_u}^2$ and $m_{H_d}^2$ parameter choices while
GMM$^{\prime}$ trades these for the more convenient weak scale parameters $\mu$ and $m_A$.)

\subsection{Results in the $m_0^{MM}$ vs. $m_{1/2}^{MM}$ plane}

A panoramic view of some of our main results is conveniently displayed in
the $m_0^{MM}$ vs. $m_{1/2}^{MM}$ plane which is then analogous to the
$m_0$ vs. $m_{1/2}$ plane of the mSUGRA/CMSSM or NUHM2,3 models.
Here, we define $m_0^{MM}=\sqrt{c_m}\alpha (m_{3/2}/16\pi^2)$ which is the
pure moduli-mediated contribution to scalar masses. The moduli-mediated
contribution to gaugino masses is correspondingly given by
$m_{1/2}^{MM}\equiv \alpha m_{3/2}/(16\pi^2)$.

\begin{figure}[H]
\begin{center}
\includegraphics[height=0.23\textheight]{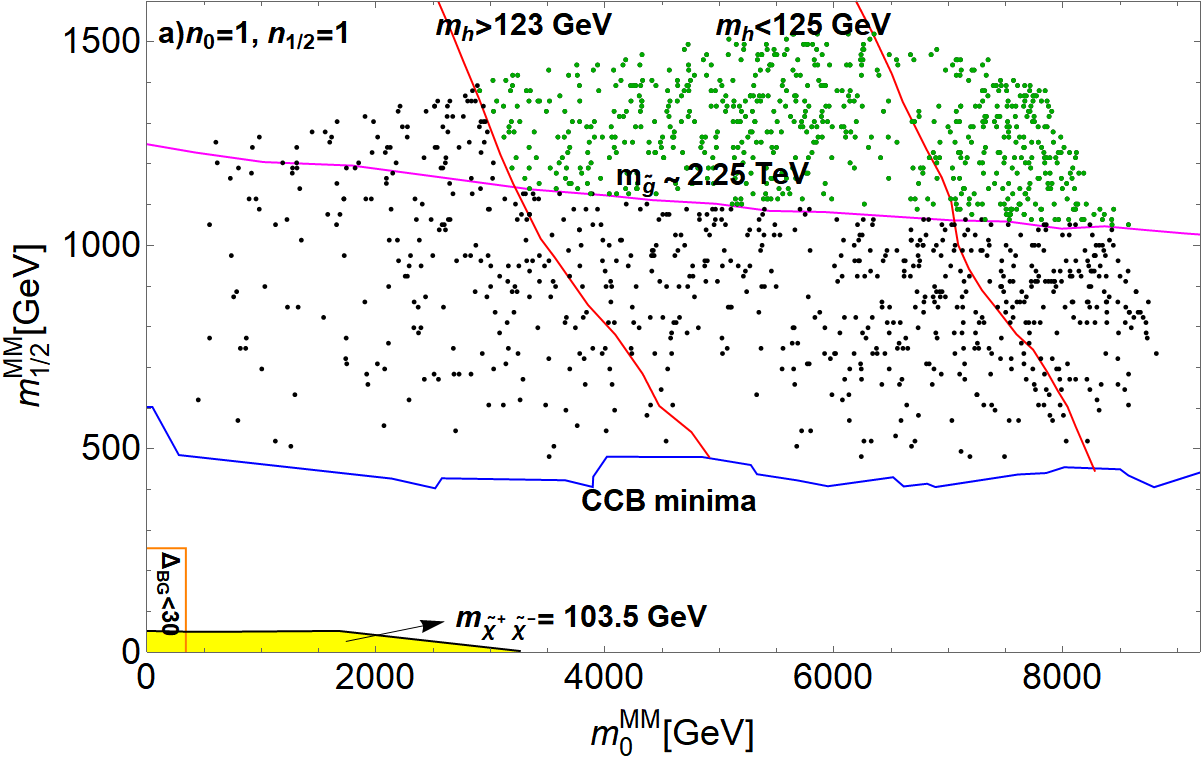}
\includegraphics[height=0.23\textheight]{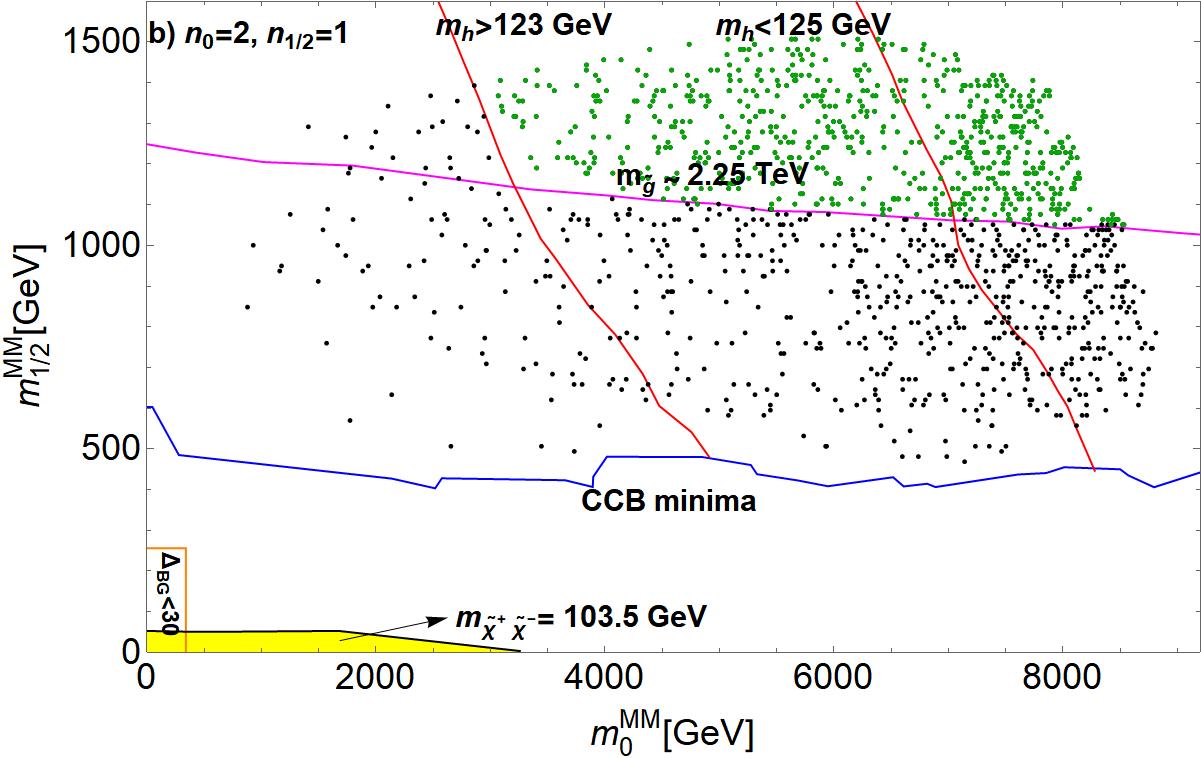}\\
\includegraphics[height=0.23\textheight]{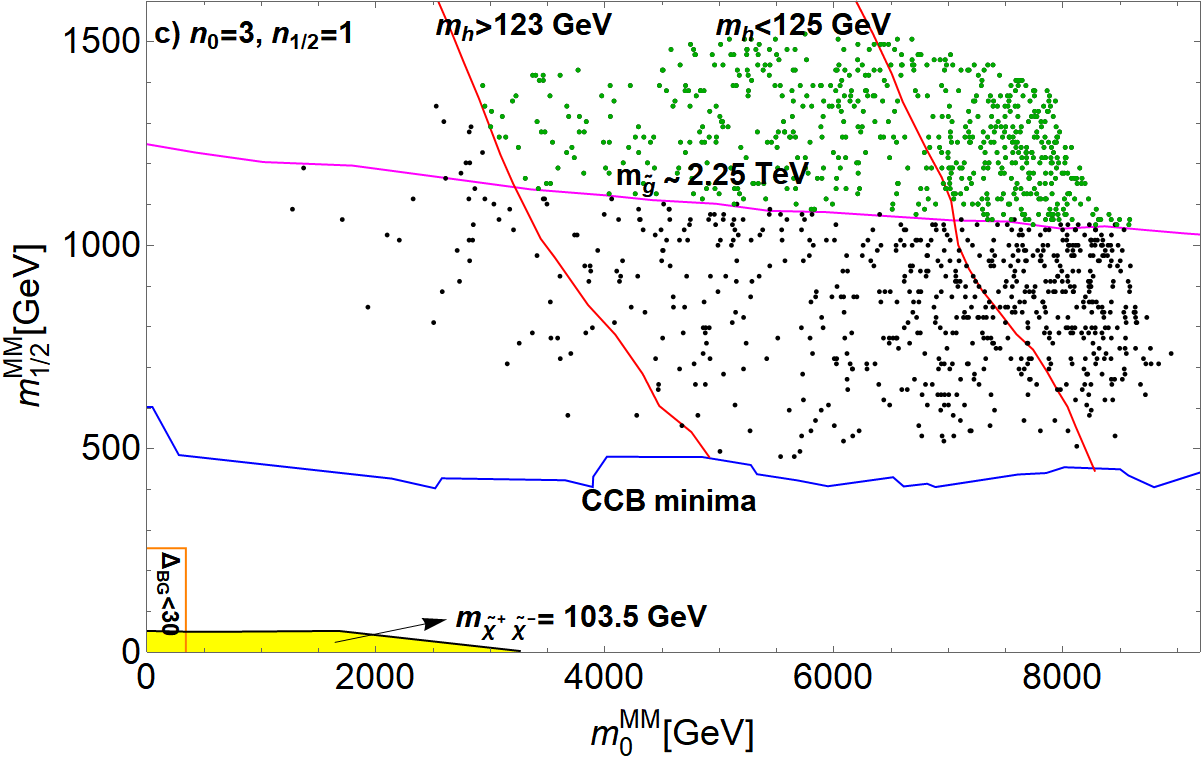}
\includegraphics[height=0.23\textheight]{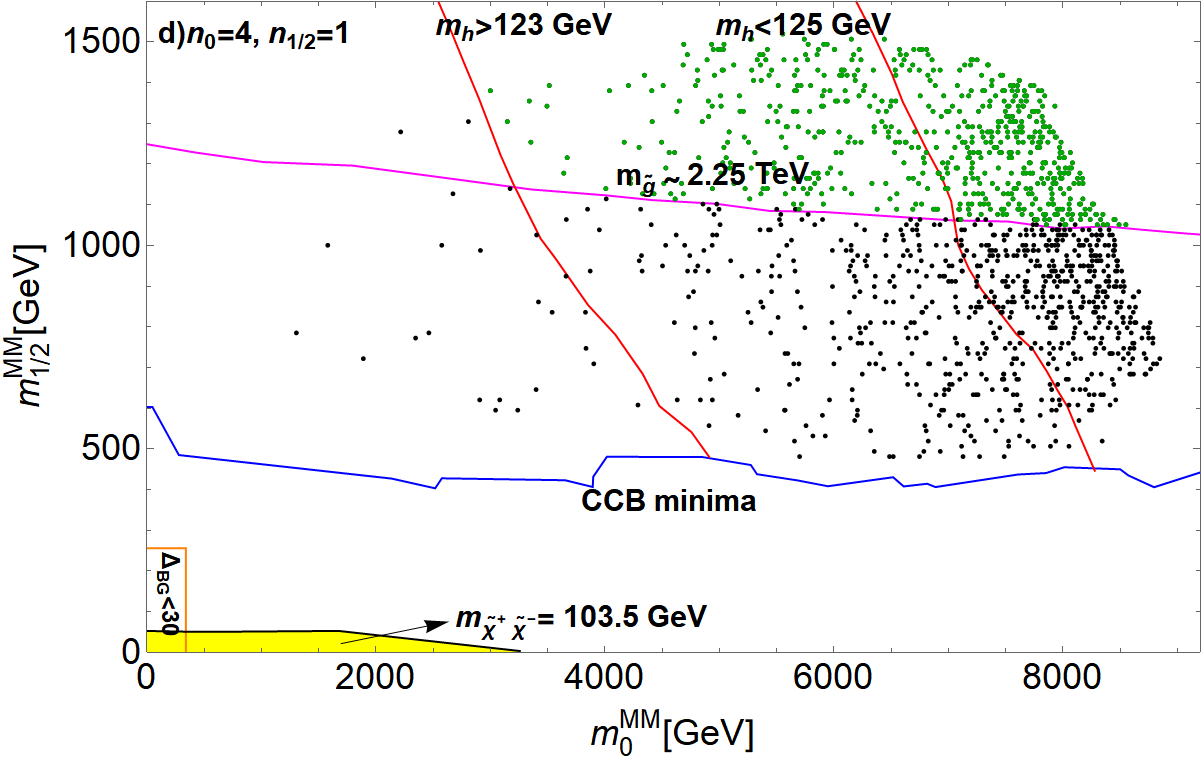}
\caption{The $m_0^{MM}$ vs. $m_{1/2}^{MM}$ plane of the GMM$^{\prime}$ model 
for a value of $n_{1/2}=1$ for all frames but with 
{\it a}) $n_0=1$, {\it b}) $n_0=2$, {\it c}) $n_0=3$ and {\it d}) $n_0=4$. 
For all frames, we take $m_{3/2}=20$ TeV, $\mu =200$ GeV, $m_A=2$ TeV, 
$\tan\beta =10$ and $a_3=1.6\sqrt{c_m}$. We require $m_Z^{PU}<4m_Z^{OU}$.
\label{fig:m0mhf}}
\end{center}
\end{figure}

In Fig. \ref{fig:m0mhf}{\it a}), we show the $m_0^{MM}$ vs. 
$m_{1/2}^{MM}$ plane for the case of an $n_{1/2}=n_0=1$ landscape draw but with 
$a_3=1.6\sqrt{c_m}$, with $c_m=c_{m3}$ and with $\tan\beta =10$, 
$m_A=2$ TeV and $\mu =200$ GeV.
The lower-left yellow region shows where $m_{\tw_1}<103.5$ GeV in violation
of LEP2 constraints. Also, the lower-left orange box shows where
$\Delta_{BG}<30$ (old naturalness calculation). The bulk of the low
$m_{1/2}$ region here leads to tachyonic top-squark soft terms owing to 
the large trilinear terms $A_0^{MM}\equiv -a_3\alpha (m_{3/2}/16\pi^2)$. 
This region is nearly flat with increasing $m_0$ mainly because the larger
we make the GUT scale top-squark squared mass soft terms, 
the larger is the cancelling correction from RG running.
For larger $m_{1/2}^{MM}$ values, then we obtain viable EW vacua since large
values of $M_3$ help to enhance top squark squared mass running to large
positive values.
The dots show the expected statistical result of scanning the landscape,
and the larger density of dots on the plot corresponds to greater
{\it stringy naturalness}. We also show the magenta contour of
$m_{\tg}=2.25$ TeV, below which is excluded by LHC gluino pair searches\cite{atlas_gl,cms_gl}.
We also show contours of $m_h=123$ and 125 GeV. The green points are
consistent with LHC sparticle search limits and the Higgs mass measurement.
From the plot, we see that much of the region of high stringy naturalness 
tends to lie safely beyond LHC sparticle search limits while at the same
time yielding a Higgs mass $m_h\simeq 125$ GeV.
While early naturalness calculations preferred low $m_0$ and $m_{1/2}$ 
regions\cite{eenz,bg,dg,ac}, we see now that stringy naturalness prefers 
the opposite\cite{Baer:2019cae}: as large as possible values of $m_0^{MM}$ and $m_{1/2}^{MM}$
subject to the (anthropic) condition that $m_{weak}^{PU}$ is within a 
factor four of our measured value (lest the atomic principle be violated).
Thus, the most stringy natural region statistically prefers a light Higgs mass
$m_h\simeq 125$ GeV with sparticles beyond LHC Run 2 reach.

In frame Fig. \ref{fig:m0mhf}{\it b}), we increase the value of $n_0$ to 2
while keeping $n_{1/2}$ fixed at 1. Likewise, in frames {\it c}) and
{\it d}), we increase $n_0$ to 3 and 4 respectively. 
The number of dots in the various frames are normalized to $\sim 1500$ so that
the relative density, indicating the relative stringy natural regions, 
can be compared on an equal footing. 
As $n_0$ increases, corresponding to more moduli fields contributing to 
SUSY breaking in the scalar sector, then the stringy natural region 
migrates towards higher values of $m_0^{MM}$ and a {\it sharpening}
of the Higgs mass prediction that $m_h\simeq 125$ GeV. 
In fact, in frame {\it d}) for $n_0=4$, there are only a few scan points 
with $m_h<123$ GeV. An anti-intuitive conclusion from our
calculations is that a 3 TeV gluino is more stringy natural than a 300 GeV gluino.  

\begin{figure}[!htbp]
\begin{center}
\includegraphics[height=0.35\textwidth]{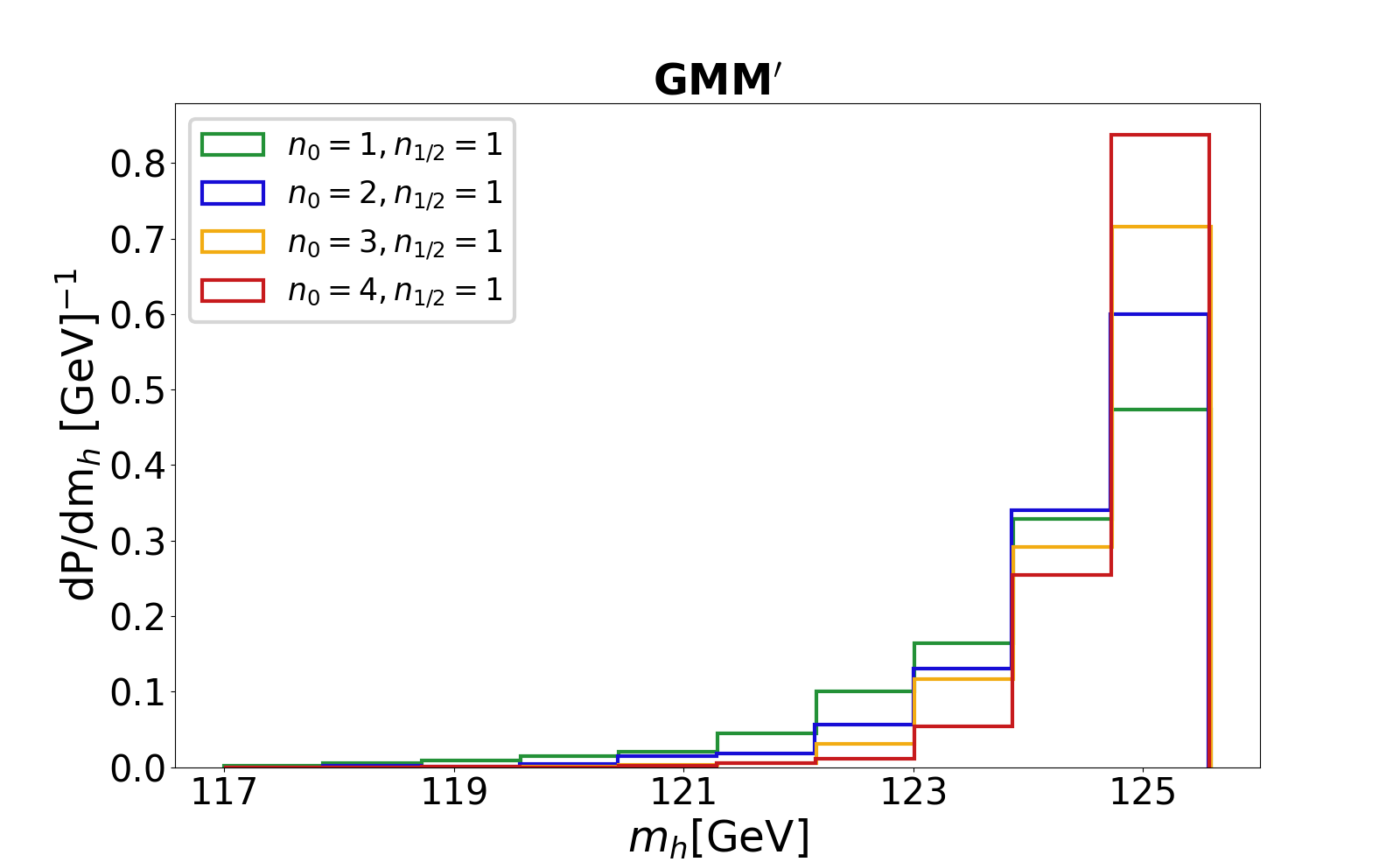}
\caption{Probability distribution $dP/dm_h$ versus $m_h$ for
$n_{1/2}=1$ and $n_0=1,\ 2,\ 3$ and $4$ 
for scans of the the GMM$^{\prime}$ model 
for $m_{3/2}=20$ TeV, $\mu =200$ GeV, $m_A=2$ TeV, 
$\tan\beta =10$ and $a_3=1.6\sqrt{c_m}$. We require $m_Z^{PU}<4m_Z^{OU}$.
\label{fig:mh}}
\end{center}
\end{figure}

In Fig.~\ref{fig:mh}, we show the histograms of Higgs mass probability
for $n_{1/2}=1$ with $n_0=1,\ 2,\ 3$ and 4. As seen from the plot, as
$n_0$ increases, the probability distribution $dP/dm_h$ does indeed
sharpen around the value of $m_h\simeq 125$ GeV.

\subsection{Parameter space scan procedure for GMM$^{\prime}$ on the landscape}

We use Isajet to scan the GMM$^{\prime}$ model parameter space as follows.
\bi
\item We select a particular value of $m_{3/2}=20$ TeV which then fixes the 
AMSB contributions to SSB terms.
\item We also fix $\mu =200$ GeV for a natural solution to the SUSY 
$\mu$ problem.
This then allows for arbitrary values of $m_Z^{PU}$ to be generated 
but disallows any possibility of fine-tuning $\mu$ to gain 
the measured value of $m_Z^{OU}$ in our universe.
\ei
Next, we will invoke Douglas' power law 
selection\cite{Douglas:2004qg,susskind,ArkaniHamed:2005yv} of moduli-mediated 
soft terms relative to AMSB contributions within the GMM$^{\prime}$ model.
Thus, for an assumed value of $n_{1/2}$ and $n_0$, we will generate
\bi
\item $\alpha^{n_{1/2}}$ with $\alpha:3-25$, corresponding to 
a power law statistical selection for moduli/dilaton-mediated 
gaugino masses $M_a$ $(a=1-3$ over the gauge groups).
\item $(a_3\alpha)^{n_0}$, a power-law statistical selection of moduli-mediated
$A$-terms, with $(a_3\alpha):3-100$,
\item $(\sqrt{c_{m3}\alpha^2})^{n_0}$ to gain a power-law statistical 
selection on third generation scalar masses $m_0(3)$, 
with $(\sqrt{c_{m3}\alpha^2}): 3-80$
\item $(\sqrt{c_{m}\alpha^2})^{n_0}$ to gain a power-law statistical 
selection on first/second generation scalar masses $m_0(1,2)$, 
with $(\sqrt{c_{m}\alpha^2}):\ \sqrt{c_{m3}\alpha^2} -320$
\item a power-law statistical selection on $m_{H_d}^2$ via $m_A^{n_0}$
with $m_A:300-10,000$ GeV.
\item a uniform selection on $\tan\beta: 3-50$.
\ei
We adopt a uniform selection on $\tan\beta$ since
this parameter is not a soft term.
Note that with this procedure-- while arbitrarily large soft terms are
statistically favored-- in fact they are all bounded from above since
once they get too big, they will lead either to non-standard EW vacua 
or else too large a value of $m_Z^{PU}$. In this way, models such as
split SUSY or high scale SUSY would be ruled out since for a 
fixed (natural) value of $\mu$ (which is not then available for fine-tuning), 
they would necessarily lead to $m_Z^{PU}\gg (2-5)m_Z^{OU}$.

{\subsection{Higgs and sparticle mass distributions for varying $n_0$}

In Fig. \ref{fig:mhgen}, we show the probability distribution for
the light Higgs mass $dP/d m_h$ vs. $m_h$ from our general landscape scans
using $n_{1/2}=1$ but with $n_0=1$ (blue) and 2 (red). Both distributions 
peak around $m_h\sim 125$ GeV, but the general scan with the harder $n_0=2$
statistical draw on scalar and trilinear soft terms is more sharply
peaked around 125 GeV than the $n_0=1$ case. 
This confirms the behavior shown previously in Fig. \ref{fig:mh} for
the more restrictive scan. 
We also generated scans with $n_0=3$ and 4, but these tend to become
very inefficient since as $n_0$ increases, one gets pushed almost always into
no EWSB or CCB minima, or minima with too large a value of $m_{weak}^{PU}$.   
\begin{figure}[H]
\begin{center}
\includegraphics[height=0.33\textheight]{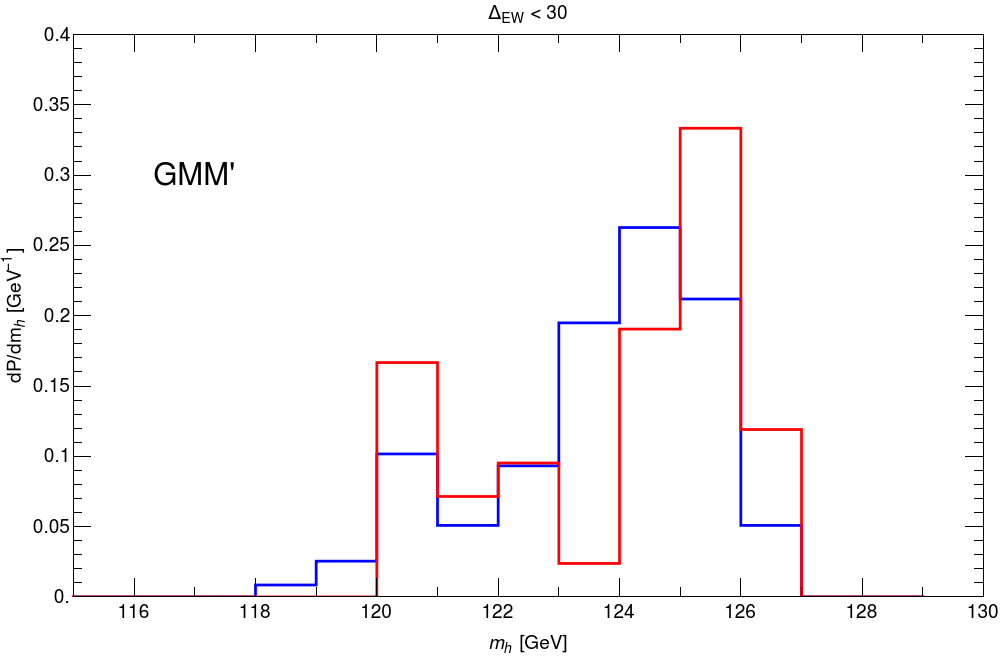}
\caption{Probability distribution for mass of light Higgs boson $m_h$
for $n_{1/2}=1$ with $n_0=1$ (blue) and $n_0=2$ (red) from 
statistical scans over the GMM$^{\prime}$ model with $m_{3/2}=20$ TeV.
\label{fig:mhgen}}
\end{center}
\end{figure}

In Fig.~\ref{fig:mass}, we show probability distributions for
{\it a}) $dP/dm_{\tg}$ vs. $m_{\tg}$, {\it b}) $dP/dm_{\tst_1}$ vs. 
$m_{\tst_1}$, {\it c}) $dP/dm_{\tst_2}$ vs. $m_{\tst_2}$ and 
{\it d}) $dP/dm_A$ vs. $m_A$.
From frame {\it a}), we see that the landscape prediction for 
$m_{\tg}$ lies between 1.5-5 TeV with a peak around 2.5 TeV for
$n_0=1$ and around 4.5 TeV for $n_0=2$.
Thus, contrary to traditional naturalness, stringy natural
predicts a gluino mass typically well above LHC mass limits.
The reach of HE-LHC with $\sqrt{s}=27$ TeV has been computed in 
Ref.~\cite{Baer:2018hpb} where the 95\% CL LHC reach with 15 ab$^{-1}$ 
was found to be $m_{\tg}\alt 6$ TeV. 
This is to be compared with the ($5\sigma$) reach of HL-LHC with 
3 ab$^{-1}$ which extends to $m_{\tg}\sim 2.8$ TeV\cite{Baer:2016wkz}. 
Thus, an energy doubling of LHC may well be required to discover 
SUSY in the $pp\to\tg\tg X$ channel.
The distributions for $m_{\tg}$ change little with varying $n_0$ 
since the gaugino mass distribution depends instead on $n_{1/2}$.
\begin{figure}[t]
  \centering
  {\includegraphics[width=.48\textwidth]{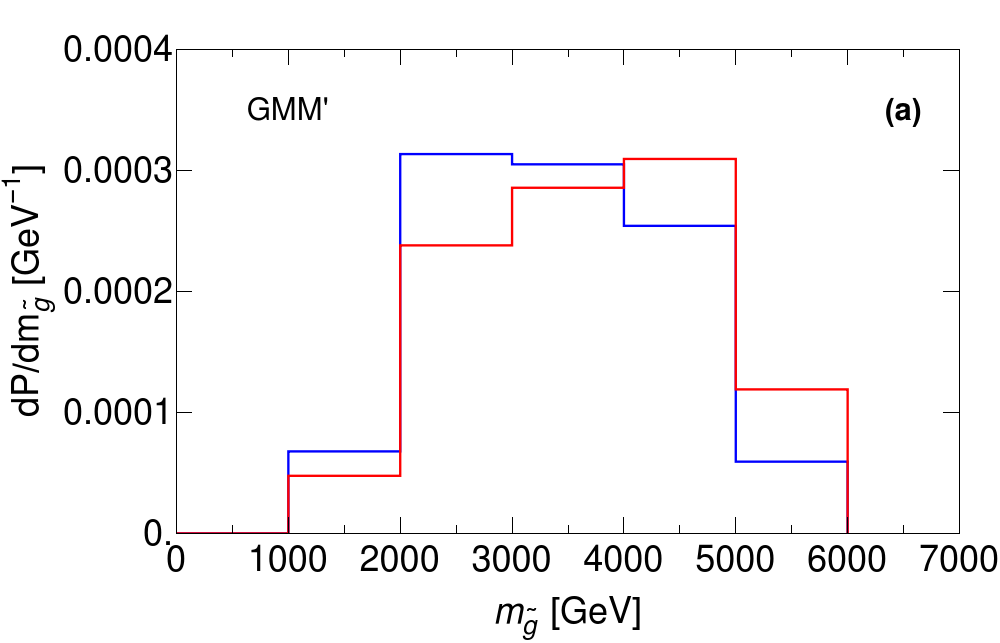}}\quad
  {\includegraphics[width=.48\textwidth]{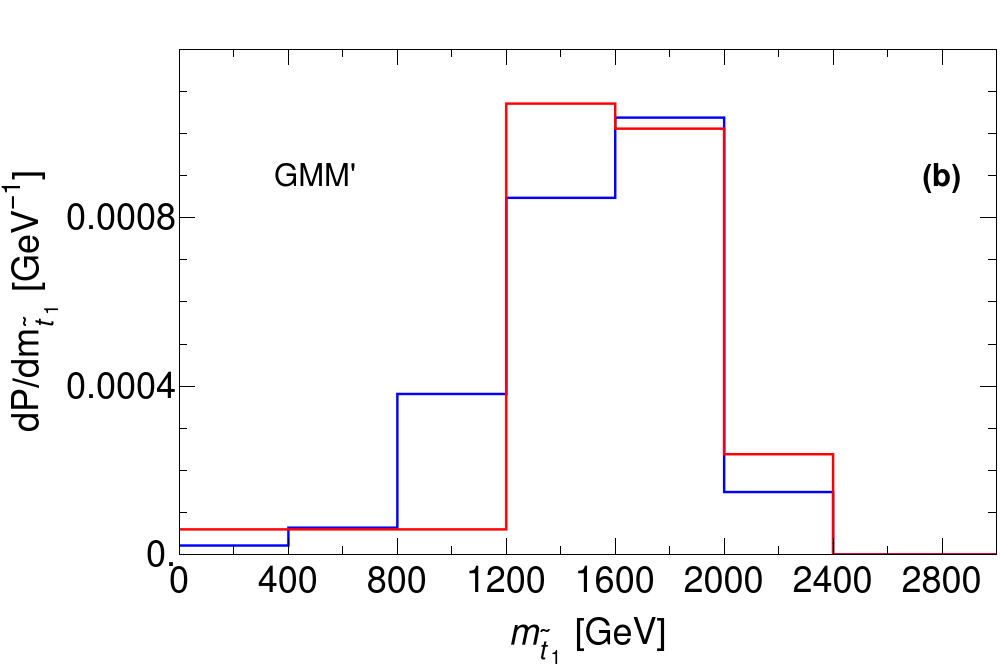}}\\ 
  {\includegraphics[width=.48\textwidth]{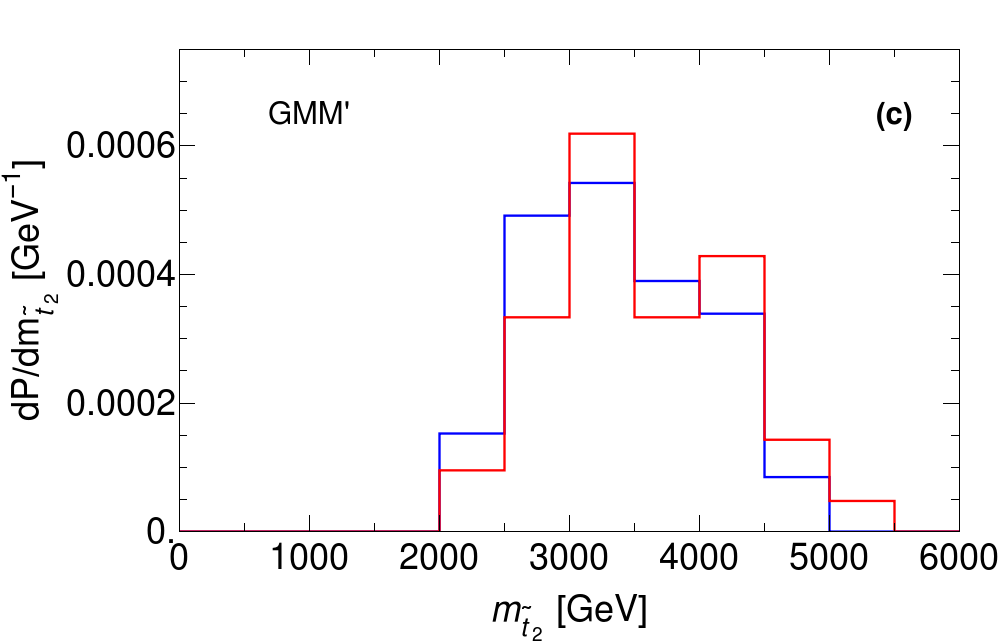}} \quad 
  {\includegraphics[width=.48\textwidth]{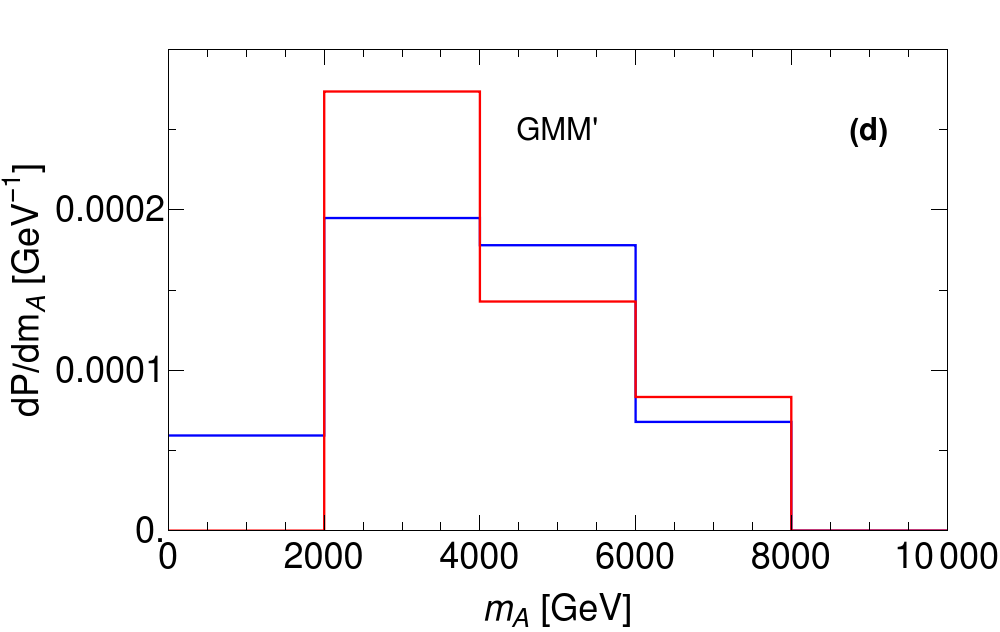}}\\
  \caption{Upper panels: Distributions in $m_{\tg}$ (left) and $m_{\tst_1}$  
(right). Lower panels: Distributions in $m_{\tst_2}$  (left) and 
$m_A$ (right). 
Here, $n_{1/2}=1$ but $n_0=1$ (blue) and $n_0=2$ (red) are from 
statistical scans over the nGMM$^{\prime}$ model with $m_{3/2}=20$ TeV. 
}  
\label{fig:mass}
\end{figure}

In frame {\it b}), the landscape probability distribution for $m_{\tst_1}$
lies between $m_{\tst_1}:1-2$ TeV with a peak probability around
$m_{\tst_1}\sim 1.5$ TeV for both cases $n_0=1$ (blue) and $n_0=2$ (red).
These distributions hardly depend on the $n_0$ value since for fixed
$\mu \sim m_{weak}$, then the largest contribution to $m_{weak}$ typically
comes from $\Sigma_u^u(\tst_{1,2})$ which sets the upper bound on $m_{\tst_1}$.  
The current limit from LHC Run 2 is that $m_{\tst_1}\agt 1.1$ TeV\cite{atlas_t1,cms_t1}. 
Thus, we see that LHC Run 2 has {\it only started} exploring the predicted 
stringy natural parameter space via stop pair production. For comparison, 
the $5\sigma$ (95\% CL) HE-LHC reach with 15 ab$^{-1}$ extends to 
stop masses of 3 (3.5) TeV. Thus, again we would require an approximate
doubling of LHC energy in order to cover the entire range of stop masses 
in landscape SUSY.

In frame {\it c}), we see the landscape prediction for $m_{\tst_2}$ lies in
the 2-5 TeV range. The reach of HL- and HE-LHC for $\tst_2$ should be 
similar to their reaches for $m_{\tst_1}$. Thus, we would expect HE-LHC
to cover only about half the expected mass range for the heavier 
top-squark $\tst_2$. The predicted statistical distribution for 
$m_{\tst_2}$ shifts to higher $m_{\tst_2}$ values for larger $n_0$
as might be expected.

In frame {\it d}), we find the distribution for $m_A$ to lie within the
$m_A\sim 1-8$ TeV range with a peak around $m_A\sim 3$ TeV for both 
$n_0=1$ and $n_0=2$. 
The upper bound on $m_A$ comes from the $m_{H_d}^2/( \tan^2\beta -1)$ term in 
Eq.~\ref{eq:mzs}: if it is too large, then $m_{weak}^{PU}$ will become too large.
From this point of view, it is not surprising that
the Higgs sector looks highly SM-like at LHC so far since there is a
decoupling of heavier Higgs particles embedded mainly in the $H_d$
multiplet while the $H_u$ multiplet is very SM-like.

In Fig. \ref{fig:muL}, we show the string landscape prediction for
first/second generation matter scalars, as typified by $m_{\tu_L}$.
From this plot, for $n_{1/2}=1$ and $n_0=1$, then we see that first/second 
generation matter scalars extend from 10-35 TeV with a peak distribution
around $m_{\tu_L}\sim 22$ TeV. The upper bound on 
first/second generation matter scalars arises not from Yukawa terms 
(tiny) or $D$-terms (which largely cancel) but from 2-loop RGE 
contributions which, if they get too large-- can drive top-squark 
soft terms to tachyonic values. As we increase $n_0$ to 2, then the
distribution in $m_{\tu_L}$ hardens even further to a peak around
$m_{\tu_L}\sim 30$ TeV. 
Both first and second generation matter scalars are pulled to a common 
upper bound since the two-loop RGE terms are flavor independent. 
This leads to the string landscape mixed quasi-degeneracy/decoupling
solution to the SUSY flavor and CP problems\cite{Baer:2019zfl}. 
In fact, in previous times model builders fought a hard battle to find
schemes which lead to universal scalar masses as a means to solve the
SUSY flavor problem. In contrast, in the string landscape picture, 
the expected non-universality of scalar masses turns out to be an asset
since the different soft terms can be drawn to sufficiently large values 
while their contributions to the weak scale remain small. 
This mechanism leads to its own mixed quasi-degeneracy/decoupling 
solution to the SUSY flavor and CP problems.\footnote{See also
Ref. \cite{Dudas:2019gxd}.}
\begin{figure}[H]
\begin{center}
\includegraphics[height=0.257\textheight]{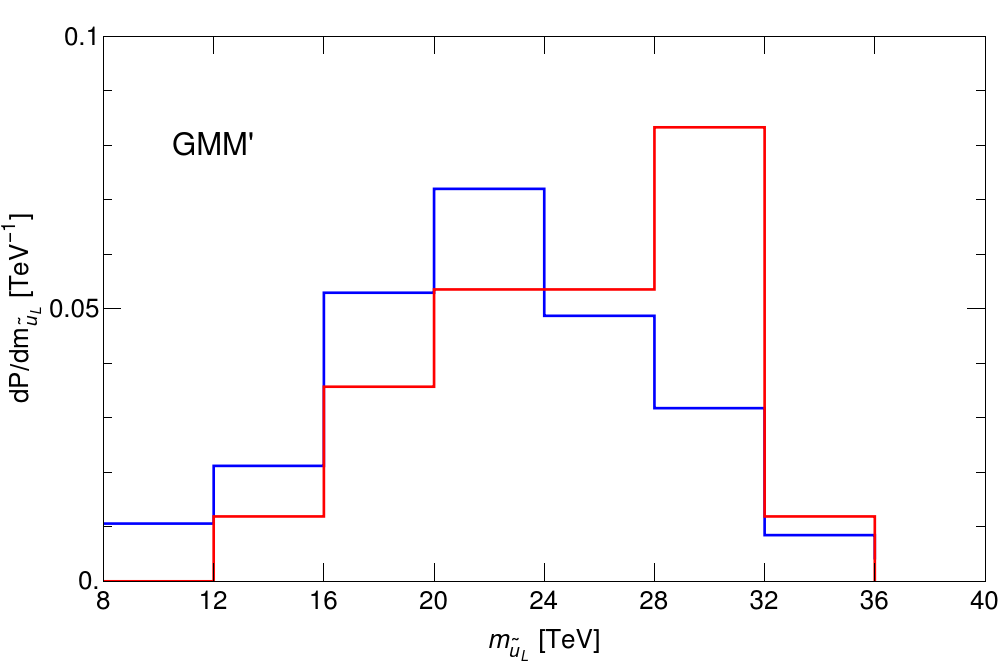}
\caption{Probability distribution $dP/dm_{\tu_L}$ vs. 
$m_{\tu_L}$ from general scan for $n_{1/2}=1$ but for 
$n_0=1$ and 2.
\label{fig:muL}}
\end{center}
\end{figure}

\section{Conclusions}
\label{sec:conclude}

In this paper, our main goal was to examine the form of soft SUSY breaking 
terms that would arise in string compactifications to a $4-d$, $N=1$ 
supergravity theory including the MSSM as the low energy EFT. 
We assumed the EFT consisted of the usual MSSM visible sector fields along
with a hidden sector of moduli fields which would serve as the arena for 
SUSY breaking. 
Using the well-known formulae for soft SUSY breaking terms in $N=1$ SUGRA, 
then we would expect the gaugino masses $m_{1/2}$, the various scalar 
masses $m_0(i)$, $m_{H_{u,d}}$ and the $A$-terms to scan independently
due to their different functional dependence on the moduli fields.

For the soft breaking scalar masses, we expect generally non-universal soft
terms due to different dependence of the K\"ahler metric on the compactified 
space. This reflects the expected {\it geography} of visible sector fields
on the compactified manifold, as emphasized by Nilles and 
Vaudrevange\cite{Nilles:2014owa}, whose conclusions were drawn 
from the context of heterotic orbifold models. 
In past times, non-universality of soft scalar masses was a thing to be 
avoided in that it could lead to dangerous flavor-violating processes.
Various contorted model-building efforts were thus made to avoid the
generic non-universality expected from realistic string compactifications.
However, in the context of the string landscape, 
{\it scalar mass non-universality turns out to be a desired property}. 
This is because the landscape likely contains a statistical draw towards 
large soft terms, especially in the scalar mass sector. 
The draw to large $m_{H_u}^2$, which stops just short of
the living dangerously feature of ``no-EWSB'', pulls $m_{H_u}^2$ to values 
associated with radiatively-driven naturalness, 
wherein large high scale soft terms are evolved via RGEs 
to natural values at the weak scale. Likewise, $A$-terms are drawn large 
enough to generate maximal mixing in the stop sector, thus minimizing the
top-squark contributions to the weak scale whilst lifting $m_h\to 125$ GeV, 
while stopping short of such large values as to generate CCB minima in the
scalar potential. Also, first/second generation scalars are drawn to 
a common upper bound in the 20-40 TeV range which leads to a mixed
quasi-degeneracy/decoupling solution to the SUSY flavor and CP problems.

We also examined the soft terms in the context of how strongly they would 
be statistically drawn to large values by the string landscape. 
In many viable string models, the tree level gauge kinetic function depends only on the
dilaton field so that a statistical pull of $m(gaugino)^{n_{1/2}}$ with
$n_{1/2}=1$ is expected. 
In contrast, the scalar masses and $A$-terms typically depend on all 
the moduli fields which would contribute to SUSY breaking, and thus a much
stronger draw of $m_{soft}^{n_0}$ with $n_0\gg 1$ may be expected.

We illustrated the consequences of these different statistical draws
in our scans over generalized mirage-mediation model GMM$^{\prime}$ parameter space
wherein comparable moduli-mediated and anomaly-mediated contributions
to soft terms arise. The cases with $n_0>n_{1/2}$ lead to predictions of
greater splitting in the SUSY particle mass spectrum with
first/second generation scalar masses $\gg$ third generation and gaugino 
masses. As $n_0$ increases relative to $n_{1/2}$, the Higgs mass 
probability distribution sharpens even more to its expected  peak 
at $m_h\sim 125$ GeV. 

What are the phenomenological consequences of the string landscape for
LHC and dark matter searches? 
Our results are summarized in Table \ref{tab:mass} from our scans over the 
GMM$^{\prime}$ model with $n_{1/2}=1$ and with $n_0=1$ or 2.
Typically, our statistical landscape approch to SUSY phenomenology predicts 
a Higgs mass $m_h\simeq 125$ GeV with sparticle masses beyond LHC reach.
Since the landscape predicts $m_{\tg}\sim 3.5\pm 2.5$ TeV and 
$m_{\tst_1}\sim 1.6\pm 0.8$ TeV, then an energy upgrade
of LHC to at least $\sqrt{s}\sim 27$ TeV may be needed for SUSY discovery
in the gluino pair or top-squark pair production channels. 
However, since the higgsino mass parameter $\mu$ is required 
not-to-far from $m_{weak}\sim 100$ GeV, it might be possible for LHC experiments 
to eke out a signal from direct higgsino pair production reactions 
such as $pp\to \tz_1\tz_2$ in the soft, opposite-sign dilepton channel\cite{Baer:2011ec}, 
perhaps in association with a hard jet radiation\cite{Baer:2014kya,Baer:2014kya2,Baer:2014kya3,Baer:2014kya4}. 
The parameter space for this SUSY discovery channel is just beginning 
to be explored\cite{Canepa:2020ntc}. 
\begin{table}[H]
\renewcommand{\arraystretch}{1.0}
\begin{center}
\begin{tabular}{|c|cc|}
\hline
mass & $n_0=1$ & $n_0=2$ \\
\hline
$m_h$ & $125^{+1}_{-4}$ GeV & $125^{+1}_{-4}$ GeV \\
\hline
 $m_{\tg}$ & $3.5\pm 2.5$ TeV & $4\pm 2$ TeV \\
\hline
$m_{\tst_1}$ & $1.6\pm 0.8$ TeV & $1.6\pm 0.8$ TeV \\
\hline
 $m_{\tst_2}$ & $3.5\pm 1.5$ TeV & $3.5\pm 1.5$ TeV \\
\hline
$m_{A}$ &  $4\pm 2$ TeV & $4\pm 2$ TeV \\
\hline
$m_{\tf}(1,2)$ & $22\pm 10$ TeV & $30^{+6}_{-18}$ TeV \\
\hline
\end{tabular}
\caption{Expected range of Higgs and sparticle masses in the
generalized mirage mediation (GMM$^{\prime}$) model from the string
landscape with $n_{1/2}=1$ but with $n_0=1$ or $n_0 =2$.
}
\label{tab:mass}
\end{center}
\end{table} 

Regarding dark matter, we would expect it to be composed of both 
SUSY DFSZ axions\cite{Bae:2013bva,Bae:2013hma} 
(which have a suppressed coupling to photons\cite{Bae:2017hlp}) along with
a smaller component ($\sim 10-20$\%) of higgsino-like WIMPs\cite{Baer:2013vpa}. 
The multi-ton noble liquid detectors now being deployed should have 
future sensitivity to the entire expected  parameter space\cite{Baer:2016ucr}, 
so we would expect a WIMP discovery should still be forthcoming in the next 5-10 years.

{\it Acknowledgements:} 

This material is based upon work supported by the U.S. Department of Energy, 
Office of Science, Office of High Energy Physics under Award Number DE-SC-0009956.


\section*{References}
\bibliography{landscape5}
\bibliographystyle{elsarticle-num}

\end{document}